\documentclass[superscriptaddress,reprint,amsmath,amssymb,footinbib,pra,aps]{revtex4-1}
\usepackage{graphicx}
\usepackage{bm}
\usepackage{color}
\usepackage{hyperref}
\usepackage{caption}
\usepackage{rotating}
\usepackage{mathtools}
\usepackage{amsmath}
\usepackage{diagbox}

\newcommand{\fig}[1]{Fig.\,\ref{#1}}
\newcommand{\eq}[1]{Eq.\ \ref{#1}}
\newcommand{\tab}[1]{Table.\ \ref{#1}}
\newcommand{\sect}[1]{Sec.\ \ref{#1}}
\newcommand{\expec}[1]{$\langle #1 \rangle$}
\newcommand{\ket}[1]{\mbox{$|#1\rangle$}}

\newcommand{\braket}[2]{\mbox{$\langle #1|#2\rangle$}}

\newcommand{\mel}[3]{\mbox{$\langle #1|#2|#3\rangle$}}

\newcommand{\SiO}{SiO$^+$}

\newcommand{\ten}[1]{$\times \text{ 10}^{#1}$}
\newcommand{\ra}{$\rightarrow$}
\newcommand{\lra}{$\leftrightarrow$}
\newcommand{\mo}{$^{-1}$ }

\newcommand{\tx}{\tilde{x}}
\newcommand{\dtx}{\Delta\tilde{x}}
\newcommand{\muto}{\mu_\text{o}}
\newcommand\Tstrut{\rule{0pt}{2.25ex}}         % = `top' strut
\begin{document}

\title{Features of Molecular Structure Beneficial for Optical Pumping}

\author{James B. Dragan}
\affiliation{Center for Fundamental Physics, Department of Physics and Astronomy, Northwestern University, Evanston, Illinois 60208, USA}
\author{Ivan O. Antonov}
\affiliation{Center for Fundamental Physics, Department of Physics and Astronomy, Northwestern University, Evanston, Illinois 60208, USA}
\affiliation{Lebedev Physical Institute, Samara Branch, 221, Novo-Sadovaya, Samara, 443011, Russian Federation}
\author{Brian C. Odom}
\email{b-odom@northwestern.edu}
\affiliation{Center for Fundamental Physics, Department of Physics and Astronomy, Northwestern University, Evanston, Illinois 60208, USA}

\date{\today}
\begin{abstract}
Fast and efficient state preparation of molecules can be accomplished by optical pumping. Molecular structure that most obviously facilitates cycling involves a strong electronic transition, with favorable vibrational branching (diagonal Franck-Condon factors, aka FCFs) and without any intervening electronic states. Here, we propose important adjustments to those criteria, based on our experience optically pumping \SiO. Specifically, the preference for no intervening electronic states should be revised, and over-reliance on FCFs can miss important features. The intervening electronic state in \SiO\text{ }is actually found to be beneficial in ground rotational state preparation, by providing a pathway for population to undergo a parity flip. This contribution demonstrates the possibility that decay through intervening states may help state preparation of non-diagonal or polyatomic molecules. We also expand upon the definition of favorable branching. In \SiO, we find that the off-diagonal FCFs fail to reflect the vibrational heating versus cooling rates. Since the branching rates are determined by transition dipole moments (TDMs) we introduce a simple model to approximate the TDMs for off-diagonal decays. We find that two terms, set primarily by the slope of the dipole moment function ($d\mu/dx$) and offset in equilibrium bond lengths ($\Delta x = r_e^g-r_e^e$), can add (subtract) to increase (decrease) the magnitude of a given TDM. Applying the model to \SiO, we find there is a fortuitous cancellation, where decay leading to vibrational excitation is reduced, causing optical cycling to lead naturally to vibrational cooling. 
\end{abstract}

\maketitle

\section{Introduction}

State-prepared molecules are of great interest in studies of ultracold chemistry \cite{Hutson2010,Quemener2012,Balakrishnan2016,Perreault2018,Liu2020,Sruthi2022a}, quantum information processing \cite{Gershenfeld1998,Demille2002,Soderberg2009,Hudson2018,Mills2020,Hudson2021}, tests of fundamental physics \cite{Chin2006,Flambaum2007,Zelevinsky2008,Demille2008,Kotochigova2009,Chin2009,Jansen2011,Kajita2014,DeMille2015,Cairncross2017,Safronova2018,Andreev2018,Kokish2018,Lim2018,Carollo2019,Kajita2020, Mitra2022} and metrology \cite{Ye2001,Schiller2014,Chou2017,Kondov2019,Alighanbari2020}. One method to achieve fast, high-fidelity state preparation is through optical pumping \cite{Viteau2008,Schneider2010,Staanum2010,Manai2012,Lien2014,Chou2017,Huang2020,Stollenwerk2020a,Antonov2021}. The challenges that arise when optically pumping molecules are from the presence of intervening vibrational and electronic states. These levels add extra decay channels and complicate the ability to attain closed optical cycling, where all excited population returns to the initial state. 

To limit experimental complexity, an obvious choice is to find molecular species whose state structure is as close to a two-level system as possible. Quantum structure that supports optical cycling involves the consideration of several core features \cite{DiRosa2004}. The main features often sought are: (1) a strong electronic transition, (2) that the transition has diagonal Franck-Condon factors (FCFs) such that the vibrational state does not tend to change and (3) no intervening electronic states where excited population can leak to, which could slow or terminate cycling. 

However, criterion (3) is called into question by our group's demonstration of optical pumping \SiO \cite{Nguyen2011,Stollenwerk2020,Antonov2021}. The SiO$^+$ molecule was chosen to demonstrate optical pumping because it contains: (1) a strong electronic $\text{X}^2\Sigma^+ \rightarrow \text{B}^2\Sigma^+$ transition and (2) diagonal FCFs \cite{Nguyen2011}. Criterion (3) is not met in \SiO\text{ }because an intervening A$^2\Pi$ state lies just 2134 cm$^{-1}$ above X$^2\Sigma^+$ \cite{Rosner1998}.  As a result, the B$^2\Sigma^+$ state can relax to either the X$^2\Sigma^+$ or A$^2\Pi$ state, such that cycling on X$\leftrightarrow$B is not closed. While it may appear to be an impediment to optical cycling, the A$^2\Pi$ state was postulated to aid in relaxation of states $\ket{\text{X},v > 2}$, which are higher in energy than $\ket{\text{A},v=0}$ \cite{Nguyen2011}. Recent results also suggested the intervening A$^2\Pi$ state assists in parity cooling \cite{Stollenwerk2020a}. 

Despite the presence of the A$^2\Pi$ state, optical pumping has been experimentally demonstrated by driving the $\text{X}^2\Sigma^+ \rightarrow \text{B}^2\Sigma^+$ transition \cite{Stollenwerk2020,Antonov2021}. These results show that it is important to reexamine the criteria of molecular features that facilitate optical pumping. In this paper, we detail exactly how the intervening electronic A$^2\Pi$ state contributes to the optical pumping process. By modeling ground rovibrational cooling of \SiO\text{, }we verify that the A state plays a critical role in parity cooling. The findings of our simulations validate a conclusion drawn from experiment \cite{Stollenwerk2020} and indicate that a discrepancy exists in the prediction for B-A state coupling between experiment and theory \cite{Qin2020}. By serving as the dominate pathway for parity flips, the intervening A$^2\Pi$ state reduces the need for an extra laser or microwave source to achieve total state preparation. These benefits from the A state show how criterion (3) can be relaxed and the ways in which intervening levels can facilitate optical pumping.

We also expand upon criterion (2) by discussing an example where favorable vibrational branching also means that vibrational cooling rates are stronger than the heating rates. Upon electronic state relaxation, off-diagonal decays --- where initial ($v_i$) and final ($v_f$) vibrational quantum numbers are not the same --- can lead to either vibrational heating ($v_f > v_i$) or cooling ($v_f < v_i$). As a first approximation, one would expect the ratio of off-diagonal FCFs, for heating and cooling decays, to predict which is the prevailing process. For example, in \SiO\text{ }analysis of the FCFs indicates that off-diagonal decays have a preference towards heating transitions by a factor of two. However, calculations of the transition dipole moments (TDMs), $\mu_{ij}$, and Einstein A-coefficients, A$_{ij} \propto \mu_{ij}^2$, which correctly predict the branching rates, indicate the opposite to be true: cooling decays occur at a $\sim$10 times higher rate. To gain insight into this discrepancy, we introduce a simple model to predict off-diagonal TDMs. The results show that certain molecular parameters can fortuitously combine to increase (decrease) the rate of vibrational cooling (heating) decays. When compared with FCF analysis alone, this model provides an easy way to better predict off-diagonal branching trends. We find that optical cycling on the X-B transition in \SiO\text{ }naturally leads to vibrational cooling. 

\section{Properties of the B-X, A-X, B-A, and X-X systems}

\begin{figure}
	\includegraphics[width=8.6cm]{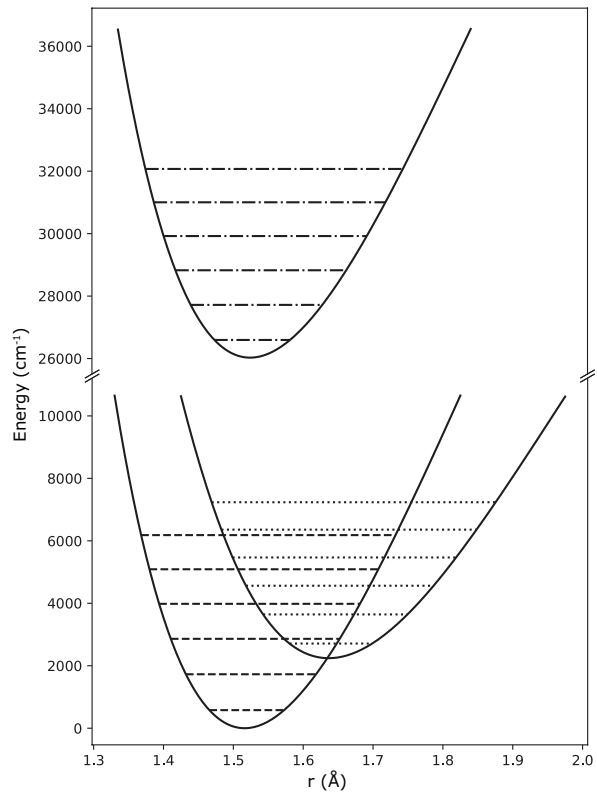}
	\caption{\footnotesize The X$^2\Sigma^+$ (dashed for vibrational states), the A$^2\Pi$ (dotted lines for vibrational states) and the  B$^2\Sigma^+$ (dot-dash lines for vibrational states) are the relevant low-lying electronic states of SiO$^+$ for optical pumping.}
	\label{fig:pecs}
\end{figure}

\begin{figure}
	\includegraphics[width=8.6cm]{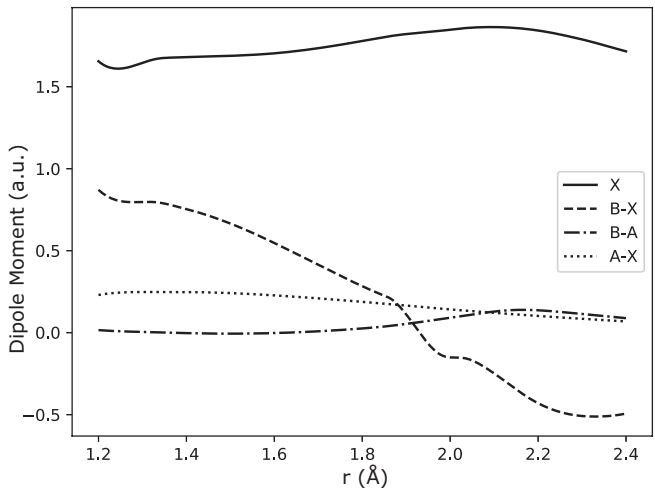}
	\caption{\footnotesize Transition dipole moments between the X$^2\Sigma^+$, A$^2\Pi$ and B$^2\Sigma^+$ electronic states of SiO$^+$ are shown. The permanent dipole moment of the X$^2\Sigma^+$ state is shown by the solid line. }
	\label{fig:dm_tdm}
\end{figure}

Optical pumping on the B-X transition of \SiO\text{ }results in populating rovibrational levels of the X$^2\Sigma^+$, A$^2\Pi$ and B$^2\Sigma^+$ electronic states. Modeling population dynamics in these states requires understanding the TDMs of the B-X, A-X and B-A electronic transitions, as well as rovibrational transitions within the X state. Potential energy curves (PECs) of the X, A and B electronic states of \SiO\text{ }obtained by Rydberg-Klein-Rees (RKR) inversion \cite{LeRoy2017} of experimental data \cite{Rosner1998} and TDM curves of B-X, B-A and A-X transitions are shown in Fig.~\ref{fig:pecs} and Fig.~\ref{fig:dm_tdm}.

The subsections below can be summarized as follows. The B-X transition at $\sim$385 nm, has the largest TDM. These two states have similar vibrational and rotational constants, as well as similar bond lengths ($r_e^X - r_e^B = -0.008$ \AA). This makes the B-X transition an ideal choice to drive for rotational pumping \cite{Nguyen2011,Stollenwerk2020a,Antonov2021}. From direct numerical integration, we find an asymmetry exists in the TDMs between B\ra X, $\Delta v$ = $\pm$1 transitions. To gain insight, we invoke the harmonic oscillator approximation and perturbation theory as a simple model to understand this behavior. The model agrees with our calculations and we find the TDMs for B-X off-diagonal vibrational branching favors a lowering of the vibrational quantum number ($\Delta v$ = -1), over a raising ($\Delta v$ = +1), by a factor of $\sim$4. This off-diagonal branching asymmetry is advantageous for optical pumping. 

The A-X decay channel also aids in pumping, as coupling between the X and A state leads to faster decay of $\ket{X,v > 1}$ states than would otherwise occur \cite{Nguyen2011}. While the B-A TDM is the weakest of the three, the TDM is not fully converged with discrepancies between the largest basis set calculation \cite{Qin2020} and previous work \cite{Cai1999,Li2019a,Li2019c}. The B-A TDM is likely even larger than predicted in Ref. \cite{Qin2020}. Discussed in Sec.\ref{howtoparityflip}, we find the decay of B\ra A\ra X is the dominate process that leads to parity flips, and a larger B-A TDM is needed to match experimental data from  Ref. \cite{Stollenwerk2020a}. %cite SI here? 

\subsection{B-X}
\label{subsec:BtoX}
The B-X system has the largest TDM of the three systems. Direct numerical integration was used to obtain FCFs and transition dipole moments, $\mu_{v',v''}$, for B-X transitions. The results are shown in \tab{tab:fcf}. As a result of nearly identical X and B equilibrium bond lengths and vibrational constants, the B-X FCFs are close to unity for $\Delta v$ = 0 transitions. These diagonal FCFs are critical for optical cycling because they allow for scattering of multiple photons in a near closed-cycle transition (e.g. $\ket{X,v=0,N=1} \leftrightarrow \ket{B,v=0,N=0} $), without exciting molecular vibrations. 

The largest off-diagonal FCFs are those for $\Delta v = \pm 1$. Interestingly, a strong asymmetry exists between the corresponding probabilities of $\Delta v = -1$ and $\Delta v = +1$ transitions. For example, the TDM $\mu_{10}$ (describing the $\ket{B,v=1} \rightarrow \ket{X,v=0}$ transition) is $\sim$ 4 times larger than $\mu_{12}$ ($\mu_{10}$ = 0.111 a.u and $\mu_{12}$ = 0.031 a.u.). Note that a simple read of the corresponding FCFs indicates the opposite: $\Delta v$ = +1 branching appears more likely than $\Delta v$ = -1 branching (2.2\% compared to 1.1\%). The pattern of TDM asymmetry continues for larger $v'$ and indicates that as the B state decays, the rate of falling down the vibrational ladder towards $v''$ = 0 is higher than climbing up it.

Such behavior can be understood within the harmonic oscillator approximation, which is valid for low vibrational levels. In this approach (see Supplementary Material \cite{SuppMaterial}) the X and B state potential energy curves are approximated as quadratic functions, with the same vibrational constant ($\omega' = \omega''$). A small shift in equilibrium bond length, $\Delta x$, in the B state is equivalent to adding a first order term to the potential energy proportional to $\Delta x$. Using perturbation theory we expand the B state vibronic wavefunctions, $\chi_{v'}^B$, as a linear combination of the X state wavefunctions, %\SI or SM?

\begin{equation}
\ket{\chi_{v'}^B} = \sum_{v'-1}^{v'+1} c_{v''} \ket{\chi_{v''}^X}.
\end{equation}

\noindent Signs of the coefficients $c_{v''}$ are dependent on the sign of $\Delta x$. In particular, 

\begin{align}
\ket{\chi_0^B} &\cong \ket{\chi_0^X} - \frac{\Delta x}{x_\text{o} \sqrt{2}} \ket{\chi_1^X} \\
\ket{\chi_1^B} &\cong \ket{\chi_1^X} + \frac{\Delta x}{x_\text{o} \sqrt{2}} \ket{\chi_0^X} - \frac{\Delta x}{x_\text{o}} \ket{\chi_2^X}.
\end{align}

\noindent Here, $x_\text{o}$ = $\sqrt{\hbar/m \omega}$ is the distance scale factor of the harmonic potential. 

If we assume the B-X TDM is a linear function of $x$ near the equilibrium bond length (see Fig. \ref{fig:dm_tdm}, where $r_e^X$ = 1.516 \AA) we can write the electric TDM as $\mu_{el}(x) \approx \mu_\text{o} + \partial_x\mu \cdot x$. Here, $\mu_\text{o}$ and $\partial_x\mu$ are zero- and first-order coefficients approximating $\mu_{el}$(x). Using the wavefunctions above, we can then approximate the functional form of the TDMs between vibrational states, $\mu_{v',v''}$, for small $v$. For off-diagonal transitions from $\ket{B,v'=1}$ to $\ket{X,v''=\{0,2\}}$ we find, 

\begin{align}
\label{eqn:mu_tdm}
\mu_{10} = \frac{x_\text{o}}{\sqrt{2}} &\Bigg(\partial_x\mu + \mu_\text{o} \frac{\Delta x}{x_\text{o}^2}\Bigg) 
\\
\mu_{12} = x_\text{o} &\Bigg(\partial_x\mu - \mu_\text{o} \frac{\Delta x}{x_\text{o}^2}\Bigg)  .
\label{eqn:mu_12}
\end{align}

\noindent In this treatment, we find conditions exist where there may be a near cancellation of the first-order term with the weighted zero-order term (when $\partial_x\mu \approx \pm \mu_\text{o} \frac{\Delta x}{x_\text{o}^2}$). If this condition is met, and both $\partial_x\mu$ and $\Delta x$ have the same sign, then $\mu_{12}$ tends towards zero and vibrational cooling ($\Delta v$ = -1) will dominate off-diagonal decays. Vibrational heating upon decay is strengthened if the two terms have different signs. In the limit where $\Delta x$ = 0, then Eq. \ref{eqn:mu_tdm} and \ref{eqn:mu_12} follow expectation that the slope of the dipole moment is the primary contributor to off-diagonal couplings, when other anharmonicities are not present or included \cite{Bernath2005}. This method provides a simple way, akin to calculating FCFs, to gain insight into which molecules are amenable to optical cycling. 

In \SiO, there is a fortuitous cancellation, as $\partial_x\mu \cong$ -1 a.u./\AA\text{ }and $\Delta x$ = -0.008 \AA\text{ }both have the same sign. As a result, the two terms add in Eq. \ref{eqn:mu_tdm} such that $\mu_{10}$ = 0.102 a.u., and cancel in Eq. \ref{eqn:mu_12}, such that $\mu_{12}$ = 0.038 a.u.. These values, computed from our simple model, compare well with the results from numerical calculations shown in Table \ref{tab:fcf}. Since the vibrational bands (1-1, 2-2, ...) are located in the same spectral region as the 0-0 band (\fig{fig:shaping}), these results indicate that any population in $v$ $>$ 0 will be pumped into $v$ = 0. 

While this model is approximate, results from direct numerical integration in Table \ref{tab:fcf} show a similar behavior.  Another important consequence for \SiO\text{ }is that the number of optical pumping cycles before off-diagonal decay is larger than was considered previously \cite{Nguyen2011}. While the $\ket{B,v=0} \rightarrow \ket{X,v=1}$ Franck-Condon factor is 1.1\%, the actual probability of decay to $\ket{X,v=1}$ is only $\approx$ 0.14\%. The higher photon budget and the net vibrational cooling discussed here would not be evident from FCF analysis alone. Interestingly, the asymmetry of transition strengths, upon B\ra X decay, between $\Delta v = +1$ and $\Delta v = -1$ continues for higher $v$'s, where the harmonic oscillator approximation becomes less valid. 

\begin{table*}
	\begin{ruledtabular}
		\begin{tabular}{c c c c}
			\diagbox{$v'$}{$v''-v'$} & 0 & +1 & -1 \\
			\hline
			0 & 98.9 / 0.642  & 1.1 / 0.024  & - \\
			1 & 96.7 / 0.620  & 2.2 / 0.031  & 1.1 / 0.111  \\
			2 & 94.5 / 0.599  & 3.3 / 0.035  & 2.2 / 0.154  \\
			3 & 92.3 / 0.578  & 4.4 / 0.037  & 3.2 / 0.187 \\
			4 & 90.1 / 0.558  & 5.5 / 0.038  & 4.2 / 0.212  \\
			5 & 87.9 / 0.537  & 6.6 / 0.038  & 5.2 / 0.234  \\
		\end{tabular}
	\end{ruledtabular}
	\caption{\footnotesize FCFs (\%) / $\mu_{v',v''}$ (a.u.) for $\Delta$ = 0, $\pm$1 transitions of the B-X system. Here, and throughout the paper, quantum numbers of the upper state are denoted by a single prime, $v'$ for example, and those of the lower state are denoted by a double prime. The TDMs in the $v''-v'$ = -1 column reflect the amplitude of vibrational cooling events upon B state decay, whereas $v''-v'$ = 1 represents that of vibrational heating. Note that a naive interpretation of the ratio of FCFs for $\Delta$v = $\pm$1 transitions would suggest a tendency toward vibrational heating rather than cooling.}
	\label{tab:fcf}
\end{table*}

\subsection{A-X}

The A-X system has the second largest TDM of the three systems, with values of  $\mu \sim$ 0.2 a.u. near the A state equilibrium bond length. Radiative decay of $\ket{A,v}$ states have lifetimes ranging from several milliseconds, for $v$ = 0, to hundreds of microseconds for states of higher $v$. The states $\ket{X,v\ge 2}$ quickly decay into lower-lying $\ket{A,v}$ states. As a result, population that enters $\ket{A,v}$ or $\ket{X,v\ge 2}$, relaxes to the $\ket{X,v=0}$ or $\ket{X,v=1}$ states on a millisecond timescale, bypassing much slower pure X-X vibrational relaxation processes. 

\subsubsection{Perturbations in the A-X system}

Rotational levels of the X and A states can interact via several coupling terms which are typically neglected in the Hamiltonian. Namely, there are spin-orbit interactions and terms in the rotational part of the Hamiltonian: homogeneous (i.e. \textbf{J}-independent) \textbf{LS} and the inhomogeneous \textbf{L}-uncoupling term \textbf{JL}. The matrix elements for these terms can be calculated based on the experimentally determined values reported in Ref. \cite{Rosner1998}. These perturbations are of several tens of cm$^{-1}$ for both the spin-orbit and \textbf{LS} terms, and $\sim$ 0.01 - 0.1 cm$^{-1}$ per \textbf{J} for \textbf{L}-uncoupling. The spin-orbit and \textbf{LS} couplings are mathematically similar and appear in different parts of the Hamiltonian --- relativistic and rotational terms respectively --- and can have different selection rules \cite{Lefebure-Brion1986}. The nominal spacing between energy levels of the A and X states is typically several hundred cm$^{-1}$. This results in small shifts which can be accounted for with some typical spin-rotation and $\Omega$-doubling terms in the Hamiltonian. However there are some closely spaced vibronic levels of A and X that result in a strong perturbation. For example, \ket{X,v=2} is nearly degenerate with \ket{A,v=0} (\fig{fig:pecs}) and this produces a large energy shift which can be seen in the \ket{X,v=2}\lra \ket{B,v=2} spectrum (\fig{fig:shaping}, top trace). These perturbative couplings also increase the probability of \ket{B,v=2} decaying to \ket{A,v=0} to 15\%. 

In addition to energy shifts, perturbations result in intensity borrowing. For example, the \ket{X,v=2} wavefunction borrows $\sim$16\% of the \ket{A,v=0} characteristics and vice versa. Therefore transitions that are normally weak (\ket{B,v=2} \ra \ket{A,v=0} and \ket{X,v=2} \ra \ket{X,v=0}) become orders of magnitude stronger \cite{Nguyen2011}. 

\subsection{B-A}
\label{BtoA}

The smallest TDM of the three systems is between the B and A states (\fig{fig:dm_tdm}). The most recent MRCI calculations \cite{Qin2020} predict an electronic TDM of \textless 0.05 a.u. near the equilibrium bond length of the A state. This results in a $\approx$ 10$^{-3}$ probability of \ket{B,v=0}\ra \ket{A,v} decay. It is worth noting that the calculated TDM is not fully converged, with the largest basis set/highest level theory calculation predicting significantly higher TDMs \cite{Qin2020} compared with the previous calculations \cite{Cai1999,Li2019a,Li2019c}.

Naively, one would think it preferable for B\ra A relaxation to be slow, since this represents a loss of population out of the optical cycling manifold. However, our modeling, discussed below, confirms that B\ra A decay is crucial for cooling the rotational degrees of freedom in \SiO. The three-step process of cycling from X$\rightarrow$B$\rightarrow$A$\rightarrow$X results in a change of total parity of the molecular wavefunction, converting odd $N$ rotational levels of the X state into even $N$ and vice versa. This provides a pathway for cooling \ket{X,v=0,N=1} to \ket{X,v=0,N=0} without the need of extra lasers or microwaves.

\subsection{Rovibrational transitions in the X state}  

Due to the $\sim $1.7 a.u.\text{ }dipole moment, the X state rotational levels can strongly interact with blackbody radiation (BBR). However, transitions between low-lying rotational states (N \textless \text{ }20) are located in the low frequency part of the blackbody spectrum, where the spectral density at room temperature is low. Subsequently, the rates for interaction with BBR are small (see \tab{tab:bbr}). Spontaneous emission rates shown in the same table are another 1-2 orders of magnitude slower. Compared with the B-X optical pumping rates, which in our experiment \cite{Stollenwerk2020a} can exceed $\sim$10$^3$ s$^{-1}$, these processes can be neglected. 

\begin{table*}
	\begin{ruledtabular}
		\begin{tabular}{c c c c c}
        J'' & J' & E (cm$^{-1}$) & A (s$^{-1}$) & B $\rho_{BBR}$ (s$^{-1}$) \\
        \hline
        0.5 & 1.5 & 1.425 & 5.70 \ten{-6} & 8.25 \ten{-4} \\
        1.5 & 2.5 & 2.870 & 5.50 \ten{-5} & 3.97 \ten{-3} \\
        2.5 & 3.5 & 4.305 & 2.00 \ten{-4} & 9.58 \ten{-3} \\
        3.5 & 4.5 & 5.740 & 4.90 \ten{-4} & 1.75 \ten{-2} \\
        4.5 & 5.5 & 7.175 & 9.70 \ten{-4} & 2.77 \ten{-2} \\
        5.5 & 6.5 & 8.609 & 1.71 \ten{-3} & 4.05 \ten{-2} \\
        6.5 & 7.5 & 10.044 & 2.74 \ten{-3} & 5.55 \ten{-2} \\
        7.5 & 8.5 & 11.478 & 4.12 \ten{-3} & 7.28 \ten{-2} \\
        8.5 & 9.5 & 12.912 & 5.91 \ten{-3} & 9.25 \ten{-2} \\
        9.5 & 10.5 & 14.346 & 8.14 \ten{-3} & 0.114 \\
        10.5 & 11.5 & 15.779 & 1.09 \ten{-2} & 0.138 \\
        11.5 & 12.5 & 17.213 & 1.42 \ten{-2} & 0.165 \\
        12.5 & 13.5 & 18.646 & 1.81 \ten{-2} & 0.193 \\
        13.5 & 14.5 & 20.078 & 2.26 \ten{-2} & 0.224 \\
        14.5 & 15.5 & 21.510 & 2.79 \ten{-2} & 0.257 \\
        15.5 & 16.5 & 22.942 & 3.39 \ten{-2} & 0.291 \\
        16.5 & 17.5 & 24.373 & 4.07 \ten{-2} & 0.328 \\
        17.5 & 18.5 & 25.804 & 4.84 \ten{-2} & 0.367 \\
        18.5 & 19.5 & 27.235 & 5.70 \ten{-2} & 0.408 \\
		\end{tabular}
	\end{ruledtabular}
	\caption{\footnotesize Transition Energies (cm$^{-1}$), Einstein A coefficients (s$^{-1}$), and rates of BBR pumping (s$^{-1}$) for \ket{X,v=0,J',e} $\rightarrow$ \ket{X,v=0,J'',e} transitions of \SiO.} 
	\label{tab:bbr}
\end{table*}

The flatness of the X-X dipole moment curve results in weak vibrational transitions. The radiative decay lifetimes of the \ket{X,v\text{ }\textgreater\text{ }0} states via vibrational ladder transitions are on the order of seconds (Fig. 2 in Ref. \cite{Nguyen2011}). However, due to the near degeneracy and perturbative coupling with \ket{A,v=0}, the \ket{X,v=2} state decay rate to \ket{X,v=0} is two orders of magnitude stronger than from \ket{X,v=1}. Rates of spontaneous emission, optical pumping by BBR and the broadband laser used in our experiments are compared in \tab{tab:op}. 

\begin{table*}
	\begin{ruledtabular}
		\begin{tabular}{c c c c c c c}
			\diagbox{$v_f$}{$v_i$} & 0 & 1 & 2 & 3 & 4 & 5 \\
			\hline
			0 & - & \textit{0.141} & \textit{34.639} & \textit{2.331} & \textit{0.413} & \textit{65.534} \\
			1 & \textbf{5.55 \ten{-4}} & - & \textit{10.185} & \textit{0.680} & \textit{0.514} & \textit{25.667} \\
			2 & \textbf{5.97 \ten{-4}} & \textbf{0.044} & - & \textit{7.226} & \textit{21.276} & \textit{18.315} \\
			3 & \textbf{1.98 \ten{-7}} & \textbf{1.39 \ten{-5}} & \textbf{0.035} & - & \textit{0.971} & \textit{3.268} \\
			4 & \hspace{4pt}\textbf{1.93 \ten{-10}} & \textbf{5.42 \ten{-8}} & \textbf{4.99 \ten{-4}} & \textbf{5.02 \ten{-3}} & - & \textit{1.505} \\
			5 & \hspace{4pt}\textbf{1.65 \ten{-10}} & \textbf{1.42 \ten{-8}} & \textbf{2.28 \ten{-6}} & \textbf{8.61 \ten{-5}} & \textbf{8.03 \ten{-3}} & - \\
		\end{tabular}
	\end{ruledtabular}
	\caption{\footnotesize Upper right triangle, italic - Einstein A coefficients (s$^{-1}$); lower left triangle, bold - BBR pumping rates (B$\ \rho_{BBR}$,s$^{-1}$) for transitions $\ket{X,v_i}$ $\rightarrow$ $\ket{X,v_f}$. Rates are averaged over the upper state's rotational levels populated at T = 300 K.}
	\label{tab:ab}
\end{table*}

\begin{table*}
	\begin{ruledtabular}
		\begin{tabular}{l l l l}
			Transition & A (s$^{-1}$) & Transition & B$\rho$ (s$^{-1}$) \\
			\hline 			
			\ket{B,v = 0} $\rightarrow$ \ket{X,v = 0} & 1.42 \ten{7} & \multicolumn{2}{c}{Optical Pumping} \Tstrut \\ \cline{3-4}
			\ket{B,v = 2} $\rightarrow$ \ket{A,v = 0} & 1.10 \ten{6} & \ket{X,v = 0, N > 0} $\leftrightarrow$ \ket{B} \ra \ket{X,v = 0} & 300 - 700 \Tstrut\\
			\ket{B,v = 1} $\rightarrow$ \ket{X,v = 0} & 4.82 \ten{5} & \ket{X,v = 1} $\leftrightarrow$ \ket{B} $\rightarrow$ \ket{X,v = 0} & 90 \\
			\ket{B,v = 1} $\rightarrow$ \ket{X,v = 2} & 3.76 \ten{4} & \ket{X,v = 0, N = 0} $\leftrightarrow$ \ket{B} $\rightarrow$ \ket{X,v = 0} & 10 \\
			\ket{B,v = 0} $\rightarrow$ \ket{X,v = 1} & 1.78 \ten{4} & \ket{X,v = 0, N > 0} $\leftrightarrow$ \ket{B} $\rightarrow$ \ket{X,v = 1} & 1.2 - 1.5 \\
			\ket{B,v = 0} $\rightarrow$ \ket{A,v = 0 - 5} & 1.31 \ten{4} & \ket{X,v = 0, N > 0} $\leftrightarrow$ \ket{B} $\rightarrow$ \ket{A} & 0.9 - 1.3 * \\
			\ket{A,v = 5} $\rightarrow$ \ket{X,v = 0 - 5} & 1.16 \ten{4} & \ket{X,v = 0, N = 0} $\leftrightarrow$ \ket{B} $\rightarrow$ \ket{X,v = 1} & 1.6 \ten{-2} \\
			\ket{X,v = 5} $\rightarrow$ \ket{A,v = 0 - 3} & 1.56 \ten{3} & \ket{X,v = 0, N = 0} $\leftrightarrow$ \ket{B} $\rightarrow$ \ket{A} & 1.2 \ten{-2} * \\
			\ket{A,v = 0} $\rightarrow$ \ket{X,v = 0 - 1} & 2.79 \ten{2} & \multicolumn{2}{c}{BBR} \\ \cline{3-4} 
			\ket{X,v = 3} $\rightarrow$ \ket{A,v = 0} & 1.52 \ten{2} & \ket{X,v = 0, N = 20} $\rightarrow$ \ket{X,v = 0} & 0.97 \Tstrut \\
			\ket{X,v = 2} $\rightarrow$ \ket{X,v = 0 - 1} & 32.5 & \ket{X,v = 1} $\rightarrow$ \ket{A,v = 0} & 0.91 \\
			\ket{X,v = 1} $\rightarrow$ \ket{X,v = 0} & 0.066 & \ket{X,v=1} $\rightarrow$ \ket{X,v=2} & 0.03 \\
			\ket{X,v = 0, N = 20} $\rightarrow$ \ket{X,v = 0, N = 19} & 0.066 & \ket{X,v = 0} $\rightarrow$ \ket{A,v = 0} & 1.6 \ten{-2} \\
			\ket{X,v = 0, N = 1} $\rightarrow$ \ket{X,v = 0, N = 0} & 5.7 \ten{-6} & \ket{X,v = 0,N = 0} $\rightarrow$ \ket{X,v = 0, N = 1} & 2.5 \ten{-3} \\
			- & - & \ket{X,v = 0,N = 1} $\rightarrow$ \ket{X,v = 0, N = 0} & 8.2 \ten{-4} \\
			- & - & \ket{X,v = 0} $\rightarrow$ \ket{X,v = 2} & 4.1 \ten{-4} \\			
			- & - & \ket{X,v = 0} $\rightarrow$ \ket{X,v = 1} & 2.8 \ten{-4} \\		 				
		\end{tabular}
	\end{ruledtabular}
	\caption{\footnotesize Comparison of spontaneous emission rates and optical pumping rates for SiO$^+$. Unless noted otherwise, rates are calculated for the lowest rotational state. *Calculated with unmodified B-A TDM value from Qin et. al. \cite{Qin2020} }  %check how to phrase this
	\label{tab:op}
\end{table*}

\section{Technical Details}

\subsection{Simulation of the SiO$^+$ Spectrum}
\label{subsec:SiOspectrum}

Spectra of the X-X, A-X, B-X and B-A systems were simulated with the PGOPHER package \cite{Western2017} using ab initio calculations and experimental data from Ref. \cite{Rosner1998}. Permanent electric dipole and transition dipole curves for the X, A, and B states were obtained from ab initio calculations \cite{Antonov2021}. Deperturbed vibrational constants of Ref. \cite{Rosner1998} were used in RKR inversion to reconstruct the potential energy curves of the X, A, and B states. The number of vibronic levels used in RKR was limited to 10 for the X state, 11 for the A state and 6 for the B state. Vibrational wavefunctions were then obtained by numerical integration of the potential energy curves and used to calculate transition dipole matrix elements between the X, A, and B states as well as perturbation matrix elements between the X and A states. The PGOPHER package was used to predict the transition energies and Einstein A coefficients for transitions between rovibrational levels of the X, A, and B states. 

\subsection{Rate Equation Model}

Optical pumping, radiative relaxation and interactions with blackbody radiation were simulated by solving a set of rate equations for state vector \textbf{N},

\begin{equation}
\frac{d \textbf{N}}{dt} = \textbf{MN}.
\label{rate}
\end{equation}

\noindent The rate coefficient matrix $\textbf{M}$ is defined as the sum of \textbf{A} + \textbf{B} + \textbf{L}, where \textbf{A}$_{ij}$ is the Einstein A coefficient connection states $i$ and $j$ ($E_j$ \textgreater $E_i$), whereas \textbf{B}$_{ij}$ (\textbf{B}$_{ji}$) and \textbf{L}$_{ij}$ (\textbf{L}$_{ji}$) are products of the Einstein B coefficients with the respective spectral densities of the blackbody field or pump laser \cite{Stollenwerk2020a} at the energy $\left| E_i - E_j \right|$. Dimensions of the equation system were defined by the number of rovibrational levels of X, A and B states, which was typically truncated at v $\le$ 5 and J $\le$ 49.5, resulting in a 2400 x 2400 matrix. The spectral energy density of the room temperature blackbody field was calculated from Planck's law. The laser spectral energy density, $\rho_L$ in (J/m$^3$/Hz), was calculated using the following equation,

\begin{align}
\notag \rho_L = &\frac{F}{100 \sigma c^2 2\pi^{3/2}} \times \\
 & \hspace{20pt} \exp{\Bigg[-\frac{(E-E_0)^2}{2 \sigma^2}\Bigg]}\frac{1}{1+e^{s(E-E_c)}}.
\label{eq:laser}
\end{align}

\begin{figure}
	\includegraphics[width=8.6cm]{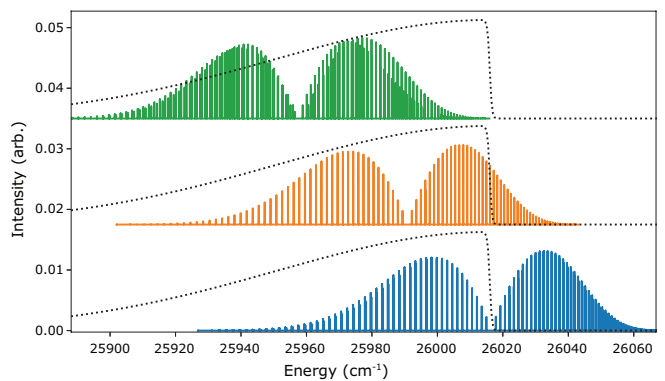}
	\caption{\footnotesize Pulse shaping mask (dotted line) used to prepare the ground rovibrational state covering the B-X 0-0 (blue, bottom trace), 1-1 (orange, middle trace), and 2-2 (green, top trace) bands. The spectral separation of the P-branch ($\Delta N =-1$) and R-branch ($\Delta N = +1$) of \SiO is evident. The orange and green bands are offset in intensity for clarity.}
	\label{fig:shaping}
\end{figure}

\noindent Here, $F$ is the laser flux, $\sigma$ is the laser's spectral bandwidth, $E_0$ is the center wavelength, $E_c$ is the pulse shaping cut-off position \cite{Stollenwerk2020a} and $s$ is the cut-off steepness. In the simulations we typically used: $F$ = 0.6 MW/m$^2$ (42 mW focused to a waist $\sim$150 $\mu$m), $\sigma$ = 65 cm$^{-1}$, $E_0$ = 26016 cm$^{-1}$ and $s$ = 3 cm. \fig{fig:shaping} shows a simulated laser spectral curve, after spectral filtering, which was used to prepare the ground rovibrational state of \SiO\text{ }\cite{Stollenwerk2020a}. Also shown are the diagonal B-X 0-0, 1-1 and 2-2 transitions. For \ket{X,v=0,N=0} state preparation, the cut-off $E_c$ was set at 26016 cm$^{-1}$, so that only the $\Delta$N = -1 transitions of the 0-0 band were pumped. The 1-1 band is pumped up to R(19.5) and higher diagonal bands are completely overlapped.

\subsection{Simulations of Ground State Cooling}

\begin{figure}
	\includegraphics[trim=2.3cm 0 0 0,clip,width=8.6cm]{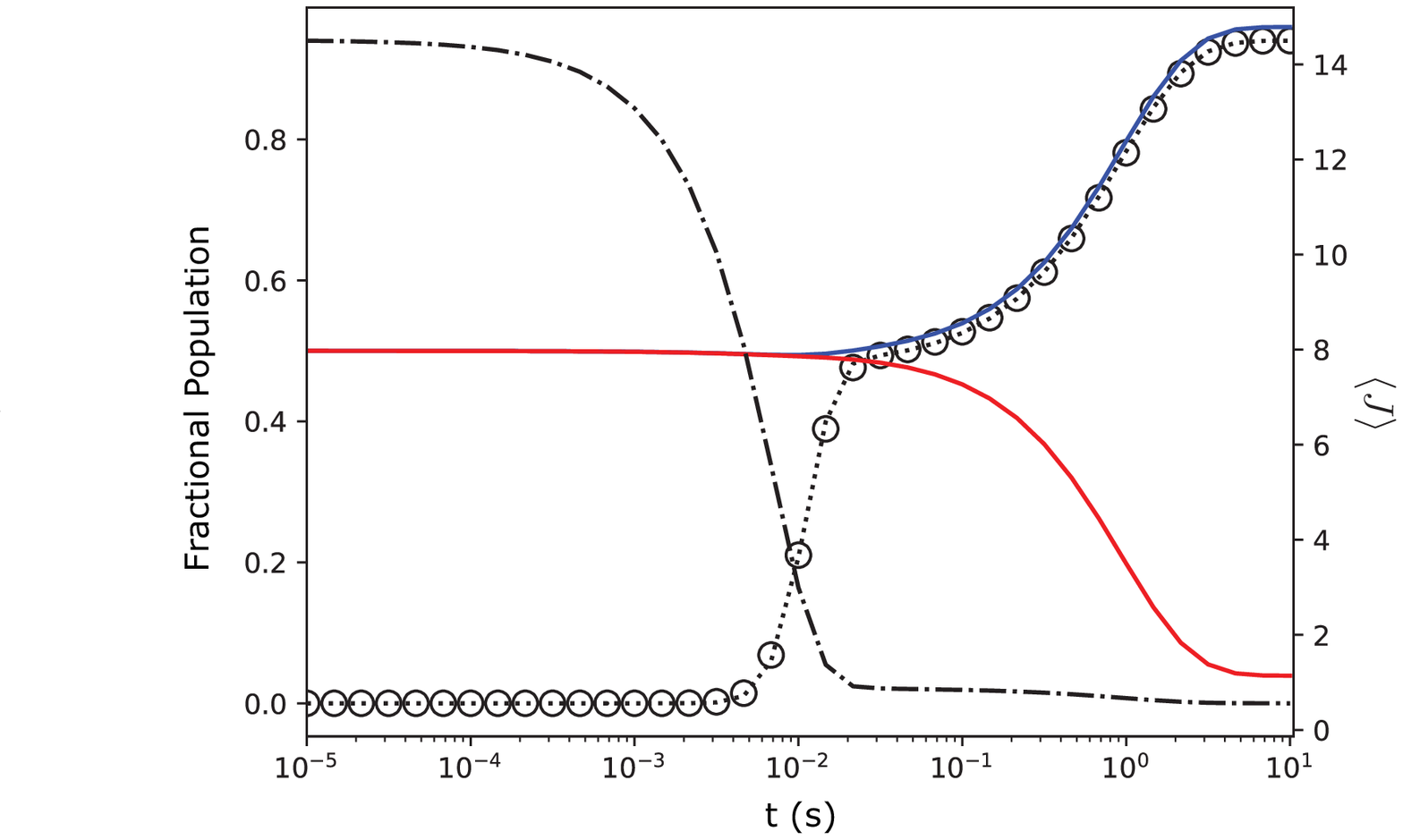}
	\caption{\footnotesize Open circles - time dependence of the simulated population in the \ket{N=0,J=0.5} state, dashed line - fit with \eq{eq:nt}, dash-dot line - average rotational level, and the red and blue solid lines show the total population in all the odd and even total parity states respectively.  %The blue line lies above the open circles because the blue line counts population in all even parity states, while the open circles are population specifically in \ket{N=0, J=0.5}.}
}
\label{fig:pumping}
\end{figure}

Cooling to the ground rovibrational state, \ket{X,v=0,N=0,J=0.5}, was simulated by setting the initial population vector \textbf{N}$_0$ and integrating \eq{rate} for a desired time range with the mask shown in \fig{fig:shaping}. Typically, it was assumed that all population at $t=0$ is equally distributed between the \ket{N=14,J=13.5}, \ket{N=14,J=14.5}, \ket{N=15,J=14.5} and \ket{N=15,J=15.5} rotational levels of \ket{X,v=0} which is consistent with the photoionization scheme used to load \SiO\text{ }at 213.96 nm \cite{Stollenwerk2019}. The time dependent population of the target state is shown in \fig{fig:pumping}.

The simulated population in the ground rotational state \ket{N=0,J=0.5} was fitted using

\begin{widetext}
\begin{equation}
n_0(t) = A_{even}\Bigg(1-e^{-\lambda t} \sum_{i=0}^{M-1} \frac{(\lambda t)^i}{i !}\Bigg) 
+ A_{odd} \Bigg[ 1 - e^{-\lambda t} 
\Bigg(\sum_{i=0}^{M} \frac{(\lambda t)^i}{i !} + \sum_{i=M+1}^{\infty} \frac{(\lambda - \kappa)^i t^i}{i !(1-\kappa/\lambda)^M} \Bigg) \Bigg]
,
\label{eq:nt}
\end{equation}
\end{widetext}

\noindent with the $A_{even}\text{, }A_{odd}\text{, }\lambda \text{ and } \kappa$ parameters determined from the fit. In the model, $M$ is equal to the number of excitation steps before reaching the lowest rotational state of a given parity. For population beginning in the rotational state $\ket{N}$, there are $M = N/2$ steps for rotational states with even parity and $M = (N-1)/2$ for rotational states with odd parity. Here, $\lambda$ is the effective rate of driving population down the rotational ladder without changing parity, i.e. reducing $N$ by 2, and $\kappa$ is an effective rate for the parity flip. These rates are convenient measures of rotational cooling at different conditions. 

\fig{fig:pumping} shows that the simulated and fitted populations of \ket{N=0,J=0.5} are in excellent agreement. Qualitatively, half of the population is initially in an even parity state and that portion reaches the ground state on a fast $\lambda \sim \text{700 s}^{-1}$ time scale. The other half in the odd parity state requires a parity flip which occurs on a much slower scale, $\kappa \sim \text{1 s}^{-1}$. This analysis is supported by the red and blue solid lines which show total population in odd and even parity states respectively, and by the dash-dot curve which shows how \expec{J} changes with time. It is clear that during the initial phase of cooling (t \textless \text{ }20 ms) both parities undergo independent rapid cooling without exchange of population and \expec{J} lowers from 14.5 to 1. At later times population with odd parity slowly converts into even and enters the ground state. Cooling is completed after 5 s with \ket{N=0, J=0.5} containing 96\% of population and \expec{J} = 0.56. 

The ground state population curve is in good agreement with the experimental data and previous simulations reported from Ref. \cite{Stollenwerk2020a}. The only disagreement being the rate of parity flip $\kappa$, measured to be 9(4) s$^{-1}$ in the experiment (labeled $\lambda_P$ in Ref. \cite{Stollenwerk2020a}). This is addressed in \sect{howtoparityflip}.   

The `fast' cooling occurs via cycles of pumping and radiative decay of the diagonal \ket{B,v=0} $\rightarrow$ \ket{X,v=0} transition. Cooling in the even parity levels (even N) proceeds via the sequence \ket{X,v=0,N=2n+2} \lra \ket{B,v=0,N=2n+1} \ra \ket{X,v=0,N=2n} until the population reaches the ground rovibrational state, the lowest level with even parity. This state is `dark', it does not interact with the pump laser because of the pulse-shaping cut-off position (see \fig{fig:shaping}). Similarly, population in the odd parity levels (odd N) proceeds via \ket{X,v=0,N=2n+1} \lra \ket{B,v=0,N=2n} \ra \ket{X,v=0,N=2n-1} until the population reaches \ket{X,v=0,N=1}. This level is not dark, and the population trapped there undergoes repeated cycling via the \ket{X,v=0,N=1} \lra \ket{B,v=0,N=0} transition until parity is flipped and it enters the dark state, \ket{X,v=0,N=0}. 

\subsection{Elucidation of the Parity Flip Mechanism}
\label{howtoparityflip}

Parity conversion requires processes with an odd number of state changes involved, e.g., \ket{X,v=0} \lra \ket{B,v=0} \ra \ket{PF} \ra \ket{X,v=0} where \ket{PF} is some intermediate `parity flip' state. This could be a vibrationally or electronically excited state, such as \ket{X,v\text{ }\textgreater\text{ }0} or \ket{A,v}. Alternatively, parity flips can occur via interaction with blackbody radiation or collisions with background gas. In an ultrahigh vacuum environment, the latter has rates 1-2 orders of magnitude too slow to explain the experimentally observed rate $\kappa$ of 9(4) s$^{-1}$ \cite{Stollenwerk2020a}. From \tab{tab:op}, the BBR rates for transitions between the relevant rotational lines occurs at an even slower rate, than collisions with background gas, and is also an insufficient explanation. However, from \tab{tab:op} we see that the decay rates from B\ra \ket{X,v\text{ }\textgreater\text{ }0} or \ket{A,v}
\ra\ket{X,v=0,N=0} are sufficiently fast to explain the observed parity flip rate $\kappa$. 

\begin{figure}
	\includegraphics[width=8.6cm]{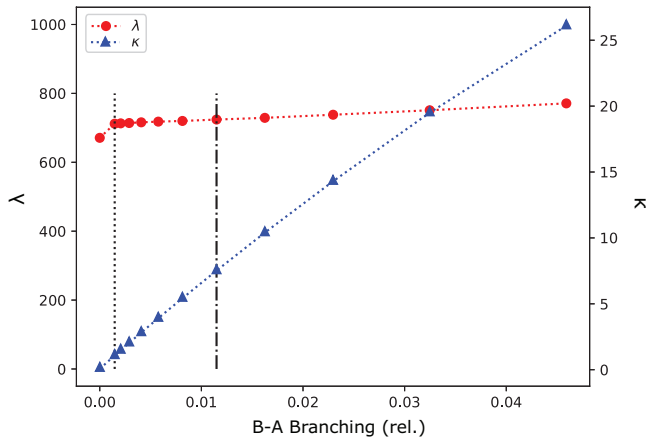}
	\caption{\footnotesize Rates for single parity cooling $\lambda$ and parity flip $\kappa$ rates as function of the B-A transition rate. Vertical lines: dashed - cooling rates from B-A branching of 0.147\% using \cite{Qin2020}, dot-dashed - cooling rates needed to fit experimental data \cite{Stollenwerk2020a} with a B-A branching of 1.15\%. The first red point is likely an artifact of the fit.}
	\label{fig:branching}
\end{figure}

To understand the contributions of the \ket{X,v\text{ }\textgreater\text{ }0} and \ket{A,v} states in the parity flipping mechanism a set of simulations was performed. Here, the B-A branching fraction, defined as the ratio of Einstein A coefficients A$_{\text{B} \rightarrow \text{A}}$/A$_{\text{B} \rightarrow \text{X}}$, was varied from 0 to 4.6\%. 
Simulated data was fit using \eq{eq:nt} and the resulting $\lambda$ and $\kappa$ cooling rates as a function of the B-A branching fraction were plotted in \fig{fig:branching}. As mentioned in \sect{BtoA}, the theory is not converged on the B-A transition dipole, and it is likely to be higher than the most recent theory work predicts \cite{Qin2020}. 

The simulated single parity cooling rate $\lambda$ weakly depends on the B-A transition rate, which has the lowest value of 671 s$^{-1}$ for zero B-A branching and the highest value of 771 s\mo for a branching of 4.6\%. The parity flip rate, $\kappa$, shows a much stronger linear dependence on B-A branching. The lowest rate of 0.0683 s\mo occurs for zero B-A branching and the highest rate is 26.1 s\mo at 4.6\%. In \fig{fig:branching} the vertical dotted line shows a branching of 0.147\% predicted for the B-A transition dipole of Ref. \cite{Qin2020}, and the dash-dot line is a branching of 1.5\%, which is needed to explain the experimentally determined $\kappa$ from Ref. \cite{Stollenwerk2020a} (labeled $\lambda_p$ there). Note this also compares well to the B-A branching measured in Ref. \cite{Stollenwerk2017}. 

The rate of parity flip in the absence of B-A branching is solely due to the $\ket{X,v>0}$ states. It is clear from \fig{fig:branching} that this contribution is too small to explain the measured rate \cite{Stollenwerk2020a}. From the mask used (\fig{fig:shaping}) and pumping timescales in \tab{tab:op}, population in $\ket{X,v=1\text{ or }2}$ is cycled to $\ket{B,v=1\text{ or }2}$, followed by off-diagonal decay to $\ket{X,v=0\text{ or }1}$, pumping excited vibrational population down to $\ket{X,v=0}$. These transitions typically result in no net change of total parity.  

This discrepancy between theory (dotted line) and experimental result (dot-dash line in \fig{fig:branching}) is a factor of $\approx$ 10. This converts into a 3.2 times larger predicted value for the B-A TDM, $\mu_{B\rightarrow A}\sim$0.064 a.u.. This matches well with the  of Ref. \cite{Stollenwerk2017}. It is plausible that the theory \cite{Qin2020} underestimates the electronic TDM magnitude in the Franck-Condon region, r $\sim$ $(r_e^A + r_e^B)/2$, of the B-A transition where it is at it's minimum of $\mu <$ 0.02 a.u. (\fig{fig:dm_tdm}).

%NOTE: why, I do not get this ! Is it bc FCF region is where the transitions are the strongest? %%%

\section{Conclusions}
Common wisdom is that molecules best for optical pumping have a strongly allowed transition between the ground and first excited electronic state, and that this transition has diagonal FCFs. Here we point out important revisions to those criteria.

Optical pumping and state preparation of molecules involves the consideration, and cooling, of parity. This is exemplified by the results from simulations, presented here, and experiments of Ref. \cite{Stollenwerk2020a}, of \SiO ground rovibrational state preparation. The overall cooling process is found to contain two timescales due to state parity. Initially, the average rotational quantum number $N$ is lowered and the parity distribution remains unchanged (see \fig{fig:pumping}) with $\ket{X,v=0,N = \{0,1\}}$ equally populated. The second timescale is the rate of pumping from $\ket{X,v=0,N=1,-}$ into $\ket{X,v=0,N=0,+}$, where the state parity is changed. 

To achieve a parity flip, microwave radiation was first proposed to transfer population \cite{Nguyen2011}. Here, we show that microwave irradiation is unnecessary, as decay through the A state is found to greatly aid in parity cooling. The three-step process, at a rate of 9(4) s$^{-1}$ \cite{Stollenwerk2020a}, involves excitation on X\ra B followed by the decay of B\ra A\ra X and results in a parity flip. This process allows for states of odd parity (odd $N$) to change to even parity. Simulations using the spectral mask, shown in \fig{fig:shaping}, show 96\% of the population is pumped to the ground rovibronic state $\ket{X,v=0,N=0,J=0.5}$ after 5 seconds. Without the A state, it takes at least 100 times longer for a parity flip to happen, extending the total rotational ground state pumping time. This critical contribution highlights an example where optical pumping benefits from the presence of an intervening electronic state.  

The merit of the A state of SiO$^+$ was first pointed out in Ref. \cite{Nguyen2011} where mixing between the A and X states increases the rate of decay from \ket{X,v > 0}\ra\ket{X,v = 0}. While the pumping into these states is hindered by the diagonal nature of the B-X transition in SiO$^+$, optical cycling in molecules with less diagonal transitions may benefit from similar intervening electronic states. In CaO$^+$, (which was recently proposed for quantum computing applications \cite{Hudson2018,Mills2020,Campbell2020}) it may be possible to pump on the highly non-diagonal X$^2\Pi$ - B$^2\Pi$ transition, provided the B state is non-dissociative. If excited vibrational X states are then populated, their decay to the ground vibrational state may be sped up by the A$^2\Sigma^+$ state. As in SiO$^+$, repeated pumping on the X\ra B transition followed by decay through the A state may allow for preparation of the ground rovibronic state of \ket{X,v\text{ }=\text{ }0} in CaO$^+$ on a short (tens of ms) time scale. 

The same logic may be applied to low-lying electronic states or even vibrational modes in polyatomic molecules. Recent work highlights the feasibility of optical cycling in linear triatomic molecules \cite{Kozyryev2016,Kozyryev2017,Augenbraun2020a,Baum2020}, symmetric and asymmetric top molecules \cite{Mitra2020,Augenbraun2020}, and other polyatomics with increasing complexity \cite{Kos2020,Dickerson2021}. Intervening electronic or vibrational states with strong transition dipoles to the ground state may assist optical pumping in a similar fashion to \SiO. For example, a bending vibrational mode in a symmetric linear triatomic molecule can facilitate radiative relaxation of a symmetric stretch mode to the ground vibrational state, which is otherwise symmetry-forbidden. A further search of molecules with low-lying electronic states, similar to \SiO\text{ }and CaO$^+$, as well as polyatomics with strong purely vibrational decay channels can expand the list of species that accommodate optical pumping.

While diagonal FCFs indicate molecules with favorable vibrational branching, the TDMs tell a more detailed story. Here, we highlight an example where FCF analysis fails to predict the reality of off-diagonal branching. In \SiO, the ratio of off-diagonal FCFs for B-X transitions suggest that vibrational heating, over cooling, is more likely. However, the rates for $\Delta v = \pm 1$ transitions (derived from TDMs) show the true propensity is for vibrational cooling. Experimentally, this implies that vibrational cooling can occur, even if only $\Delta v$ = 0 transitions are being driven. This is of great importance to optical pumping schemes, as the number of cycles before vibrational heating occurs is larger than what FCFs alone predict. 

\begin{acknowledgements}
J.B.D., I.O.A. and B.C.O. were supported by NSF grant No. PHY - 1806861. 
\end{acknowledgements}

\bibliography{pumpingv2}

\newpage
\text{ }
\vspace{20cm}
\onecolumngrid
\section*{Supplementary Material}

\twocolumngrid

\section*{Harmonic Oscillator Approximation to Vibrational Transitions in Molecules}

Suppose we have a ground state wavefunction described by a harmonic oscillator Hamiltonian. This is a valid assumption for low vibrational levels of real diatomic molecules with deeply bound electronic states. In doing so we write,

\begin{equation}
	H_g = \frac{p^2}{2m} + \frac{1}{2}m\omega^2 x^2 \text{ ,}
\end{equation}

\noindent or in dimensionless form,

\begin{equation}
	\frac{H_g}{E_\text{o}} = \frac{1}{2}\Bigg(\frac{p}{p_\text{o}}\Bigg)^2 + \frac{1}{2}\Bigg(\frac{x}{x_\text{o}}\Bigg)^2 = \frac{1}{2}\tilde{p}^2 + \frac{1}{2}\tilde{x}^2 \text{ ,}
\end{equation}

\noindent where $E_\text{o}$ = $\hbar\omega$, $p_\text{o} = \sqrt{\hbar m \omega}$, and $x_\text{o} = \sqrt{\hbar/m\omega}$ are the energy, distance and momentum scales and $\tilde{x}$ ($\tilde{p}$) is the dimensionless position (momentum) operator. Suppose the excited state has a similar potential to the ground state ($\omega = \omega'' \simeq \omega'$) and only differs in equilibrium bond length by $\Delta x$ = $x_e'' - x_e' \ll x_\text{o}$. In dimensionless form we write $\Delta\tilde{x} = \frac{\Delta x}{x_\text{o}}$. Here, we use the typical convention of molecular emission where parameters of the ground state are denoted by a double prime ($x_e''$), and those of the excited state by a single prime ($x_e'$). We then can write down the dimensionless excited state Hamiltonian as, 

\begin{equation}
	\frac{H_e}{E_\text{o}} = \frac{1}{2}\tilde{p}^2 + \frac{1}{2}\big(\tx+\dtx\big)^2 \cong \frac{H_g}{E_\text{o}} + \tx \cdot \dtx \text{ .}
\end{equation}

In the above equation, the second term can be treated as a perturbation that couples adjacent vibrational levels of harmonic oscillator states, because $\tx = \frac{1}{\sqrt{2}}(\hat{a} + \hat{a}^\dagger)$. Using perturbation theory to first order, vibrational wavefunctions of the excited state, $\ket{\chi_{v'}'}$, can be expressed in terms of those of the ground state, $\ket{\chi_{v''}''}$. For $v$ = 0,

\begin{equation}
	\ket{\chi_0'} \cong \ket{\chi_0''} - b\ket{ \chi_1''} ,
\end{equation}

\noindent where $b$ is given by,

\begin{equation}
	b = \mel{\chi_0''}{\tx\cdot \dtx}{\chi_1''} = \frac{\Delta x}{x_\text{o} \sqrt{2}} .
\end{equation}

\noindent Similarly, for $v$ = 1,

\begin{equation}
	\ket{\chi_1'} \cong \ket{\chi_1''} + b\ket{\chi_0''} - c\ket{\chi_2''} ,
\end{equation}

\noindent where $c$ is,

\begin{equation}
	c = \mel{\chi_1''}{\tx \cdot \dtx }{\chi_2''} = \frac{\Delta x}{x_\text{o}} .
\end{equation}

The transition dipole moment ($\mu_{v',v''}$) for off-diagonal vibrational decay that leads to vibrational cooling, namely $\ket{e,v'} $\ra$ \ket{g,v''=v'-1}$, can be calculated by,

\begin{equation}
	\mu_{v',v'-1} = \mel{\chi_{v'-1}''}{\mu_{el}}{\chi_{v'}'} .
	\label{eq:baremu}
\end{equation}

\noindent Here, the electronic transition dipole moment operator can be expanded around the equilibrium bond length to the first order of $\tx$ as,

\begin{align}
	\mu_{el}(x) &\approx \mu_\text{o} +  \partial_x \mu \cdot x  \\
	\notag &= \mu_\text{o} + \partial_x \mu \cdot \tilde{x} x_\text{o}
\end{align}

\noindent where $\mu_\text{o}$ [a.u.] and $\partial_x \mu$ [a.u./\AA] are linear fit coefficients of the dipole moment function given by $\mu_{el}(x-x_e) $, where $x_e$ is the equilibrium bond length of the harmonic potential. Then, for $v'$ = 1, Eq. \ref{eq:baremu} becomes,

\begin{equation}
	\mu_{10} = \mu_\text{o} \braket{\chi_0''}{\chi_1'} + \partial_x \mu \cdot x_\text{o} \mel{\chi''_0}{\tx}{\chi_1'}
\end{equation}

\noindent Substituting in the expressions for $\ket{\chi_1'}$, using the orthonormality of harmonic oscillator wavefunctions and writing $\tx$ in terms of the ladder operators, we find

\begin{align}
	\notag	\mu_{10} &= \mu_\text{o} \Big(\braket{\chi_0''}{\chi_1''} + b \braket{\chi_0''}{\chi_0''} - c \braket{\chi_0''}{\chi_2''} \Big) \\
	\notag &\hspace{10pt} + \partial_x \mu \cdot x_\text{o} \Big(\mel{\chi_0''}{\tx}{\chi_1''} + b \mel{\chi_0''}{\tx}{\chi_0''} - c \mel{\chi_0''}{\tx}{\chi_2''} \Big) \\
	\notag	&= \mu_\text{o} b + \partial_x \mu \cdot x_\text{o} \mel{\chi_0''}{\tx}{\chi_1''} \\
	&= \frac{1}{\sqrt{2}} \Bigg(\partial_x \mu \cdot x_\text{o} + \mu_\text{o} \frac{\Delta x}{x_\text{o}} \Bigg) \\
	&= \frac{x_\text{o}}{\sqrt{2}} \Bigg(\partial_x \mu + \mu_\text{o} \frac{\Delta x}{x_\text{o}^2} \Bigg) 
	\label{eq:mu10}
\end{align}

\noindent Additionally, $\mu_{12}$ one can show,

\begin{align}
	\notag \mu_{12} &= \mu_\text{o} \Big(\braket{\chi_2''}{\chi_1''} + b \braket{\chi_2''}{\chi_0''} - c \braket{\chi_2''}{\chi_2''} \Big)\\
	\notag &\hspace{10pt} + \partial_x \mu \cdot x_\text{o} \Big(\mel{\chi_2''}{\tx}{\chi_1''} + b \mel{\chi_2''}{\tx}{\chi_0''} - c \mel{\chi_2''}{\tx}{\chi_2''} \Big) \\
	\notag &= -\muto c + \partial_x \mu \cdot x_\text{o} \mel{\chi_2''}{\tx}{\chi_1''} \\
	&= \Bigg(\partial_x \mu \cdot x_\text{o} - \mu_\text{o} \frac{\Delta x}{x_\text{o}} \Bigg)\\
	&= x_\text{o} \Bigg(\partial_x \mu - \muto \frac{\Delta x}{x_\text{o}^2}  \Bigg) 
	\label{eq:mu12}
\end{align}

\noindent In terms of off-diagonal branching, one may also be concerned with $\mu_{01}$. We find,

\begin{align}
	\notag \mu_{01} &= \muto \Big(\braket{\chi_1''}{\chi_0''} - b \braket{\chi_1''}{\chi_1''}\Big) \\
	& \vspace{10pt} + \partial_x \mu \cdot x_\text{o} \Big(\mel{\chi_1''}{\tx}{\chi_0''} - b\mel{\chi_1''}{\tx}{\chi_1''}\Big) \\
	\notag &= - \muto b \braket{\chi_1''}{\chi_1''} + \partial_x \mu \cdot x_\text{o} \mel{\chi_1''}{\tx}{\chi_0''} \\
	&= \frac{x_\text{o}}{\sqrt{2}} \Bigg(\partial_x \mu - \muto \frac{\Delta x}{x_\text{o}^2}\Bigg)
	\label{eq:mu01}
\end{align}

\noindent In general,

\begin{subequations}
	\begin{align}
		\mu_{v,v-1} &= x_\text{o} \sqrt{\frac{v}{2}} \Bigg(\partial_x \mu + \frac{\muto \Delta x}{x_\text{o}^2}\Bigg) \label{eq:vm1} \tag{14}\\
		\mu_{v,v+1} &= x_\text{o} \sqrt{\frac{v+1}{2}} \Bigg(\partial_x \mu - \frac{\muto \Delta x}{x_\text{o}^2}\Bigg) \label{eq:vp1} \tag{15} 
	\end{align}
\end{subequations}

It is easy to see from Eq. \ref{eq:vm1} and Eq. \ref{eq:vp1} that the magnitude of the parentheses term is zero if $\partial_x \mu = \pm \mu_\text{o} \frac{\Delta x}{x_\text{o}^2}$. The sign in the formula is determined by signs of $\Delta x$ and $\partial_x \mu$. The former represents the difference in the equilibrium bond lengths between lower and upper electronic state, and the latter represents the slope of the electronic transition dipole near the equilibrium bond length. If $\Delta x$ and $\partial_x \mu$ have the same sign, either positive or negative, then $\mu_{v',v'+1}$ \ra 0. If the signs of $\Delta x$ and $\partial_x \mu$ are opposite, then $\mu_{v',v'-1}$ \ra 0. 

Even if the exact cancellations do not hold, corresponding transition moments are still reduced to a non-zero value compared to other off-diagonal decay. Therefore, strong discrepancies may exist between $\Delta v$ = $\pm$1 transition strengths, between $\mu_{10}$ and $\mu_{12}$ for example. In the case of the X and B states of \SiO, $\Delta x$ = -0.008 \AA, $x_\text{o}$ = 0.053 \AA, $\mu_\text{o} \cong$ 0.6 a.u. and $\partial_x \mu \cong$ -1 a.u./\AA . This results in a near cancellation of the terms in Eq. \ref{eq:vp1} such that $\mu_{12}$ = 0.037 a.u. and $\mu_{01}$ = 0.027 a.u. respectively. In contrast, these two terms add constructively in Eq. \ref{eq:vm1} giving $|\mu_{10}|$ = 0.102 a.u. This asymmetry means that decay from the B state leads to overall vibrational relaxation. This is beneficial for optical pumping schemes because a) vibrational cooling occurs when pumping the 1-1 diagonal transition which is spectrally close to the 0-0 transition and b) the total number of cycles before vibrational excitation occurs is increased (i.e. more scattering events before undesired branching).

\end{document}